%% file: 0_main.tex
\documentclass[referee]{raa_rb}
\usepackage[utf8]{inputenc}
\usepackage{graphicx}
\usepackage[pagebackref=True]{hyperref}
\usepackage{amsmath, amssymb}
\usepackage{multirow,rotating}
\usepackage{relsize}
\usepackage{array}
\usepackage{caption,adjustbox}
\usepackage{natbib}
\usepackage{changepage}

\newcommand{\lcdm}{$\Lambda$CDM }

\providecommand{\keywords}[1]{\textbf{Key words:} #1}
\newcommand\Tstrut{\rule{0pt}{2.8ex}}
\newcommand\BTstrut{\rule{0pt}{4ex}}
\newcommand\Bstrut{\rule[-1.5ex]{0pt}{0pt}}
\renewcommand{\cite}{\citep}

\newcolumntype{M}[1]{>{\centering\arraybackslash}m{#1}}
\newcommand{\mhref}[3][blue]{\href{#2}{\color{#1}{#3}}}%

\begin{document}
\title{Revisiting the epoch of cosmic acceleration}
\volnopage{Vol.0 (20xx) No.0, 000--000}
\setcounter{page}{1}

\author{David Dahiya\thanks{E-Mail:\mhref{mailto:davddahiya@gmail.com}{davddahiya@gmail.com}} and Deepak Jain\thanks{E-Mail:\mhref{mailto:djain@ddu.du.ac.in}{djain@ddu.du.ac.in}} }

\institute{$^*$Technische Universität Dresden, 01069 Dresden, Germany\\
$^{**}$Deen Dayal Upadhyaya College, University of Delhi, Dwarka, New Delhi, India\\  {\small Received 20xx month day; accepted 20xx month day}}
\abstract{
We revisit the epoch of cosmic speed-up characterized by the redshift of transition from a decelerated to an accelerated phase. This redshift is termed the transition redshift ($z_t$). We use the spatially Flat and Non-Flat variants of the most common \lcdm and XCDM models to put constraints on the transition redshift along with the other model parameters. The data for this analysis comes from the recent and updated Pantheon+ Supernova (SNe) dataset and the Hubble parameter measurements obtained from Cosmic Chronometers (CC). We consider both datasets with their respective covariance matrices incorporating all statistical and systematic uncertainties. We observe that using the combined datasets of $H(z)$ and SNe, the best fit value of transition redshift lies in the range $0.61 < z_t < 0.82$ for all four dark energy models. Incidentally, we observe a positive curvature for the Non-Flat models, correlations between several model parameters and a strong degeneracy between the curvature and the equation of state parameter.
}

\authorrunning{David Dahiya \& Deepak Jain }            
\titlerunning{Revisiting the Epoch of Cosmic Acceleration}

\maketitle

\input{1_iandm}
\input{2_data}
\include{3_results}

\include{ms2022-0477table2}
\include{4_conclusion}
\input{ms2022-0477table3}

\section*{Acknowledgements}
One of the author (David Dahiya) thanks the Principal, Deen Dayal Upadhyaya College, for providing the facilities where part of the work was done.

\bibliographystyle{raa}
\bibliography{0_main}

\end{document}

%% file: 1_iandm.tex
\section{Introduction}
The 1998 study of very distant supernovae provided irrefutable proof that, at present, the universe is undergoing an accelerated expansion.\cite{riess_1998,Perlmutter_1999}. Through high-redshift supernovae, it was established that the early universe was dominated by non-relativistic matter, which supports a decelerating expansion of the universe. Thus, it was apparent that, at a certain epoch, the expansion of the universe shifted from a decelerating phase to an accelerated one. 
This epoch is characterized by the transition redshift and denoted by the parameter $z_t$. It is suggested that $z_t$  may be a new fundamental cosmological parameter (along with $H_0$ and $q_0$) that aids in understanding the evolution of cosmic expansion \cite{lima,mel}.

In recent years, with the influx of new data, several model-independent and model-dependent approaches have been formulated to constrain the transition redshift and other parameters. The model-independent approach does not make any assumptions about the composition of the universe or the theory of gravitation other than assuming a metric structure. This approach involves  parameterizations and reconstructions of different kinematic variables, including the Hubble parameter $H(z)$, the deceleration parameter $q(z)$, and the equation of state parameter $\omega(z)$ 
 in a model independent way \cite{abdulla,cam2020,Seikel2012}. 
For instance, Rani et al. (2015) used three different parameterizations of the deceleration parameter and a local regression method to extrapolate the Hubble parameter and obtained a $z_t \in [0.60,0.98]$ \cite{Rani_2015}. Similarly, Jesus et al. (2018) measured a $z_t \in [0.806,0.973]$ using different polynomial parameterizations of the comoving distance $D_C(z)$, $H(z)$ and $q(z)$ \cite{Jesus_2018}. On the other hand, \cite{Jesus_2020} used Gaussian process to reconstruct $H(z)$ and the luminosity distance, $D_L(z)$. They obtained  transition redshift as 0.59  and 0.683 for the two reconstructions respectively. For a similar reconstruction of H(z), Toribio and Fabris (2020) obtained a $z_t\sim0.7$ \cite{Toribio_2021}. Capozziello et al. (2022) measured a $z_t \in [0.473,1.183]$ after performing a more recent reconstruction of H(z) and q(z) using SNe and Hubble data \cite{capo2022}. More methods and parameterization for obtaining $z_t$ can be found in \cite{Kumar2022,Koussour2022,muccino2022,Cunha2008,Yu2018,capo2014}.

On the other hand, the model-dependent approach, though relatively simpler, gives a much deeper intuition about the evolution of the universe and its constituents. Current observations strongly favour a universe dominated by a cosmic fluid (dark energy) with negative pressure and constant energy density. This is the standard \lcdm model of cosmology which can propel the accelerated expansion of the universe \cite{Carroll2001}. Unfortunately, there are still some inconsistencies that the model fails to address. Specifically, the fine-tuning and the coincidence problem \cite{lcdmproblem,darkenergy}. Therefore, alternate dark energy models such as the XCDM, phantom, quintessence, GCG (generalized chaplygin gas), MCG (modified chaplygin gas) etc. were considered. For example: Melchiorri et al. (2007) used MCMC methods to constrain the parameters in the \lcdm and other modified Dark energy models. The models iterated through different theoretical assumptions and parameterizations and found a $z_t \in [0.32,0.48]$ \cite{mel}. Farooq et al. (2016) used Likelihood maximization technique on three different spatially flat and non-flat models (\lcdm, XCDM, $\phi$CDM) with Hubble data from BAO and Cosmic Chronometers. Using different priors on $H_0$ they found the value of $z_t \in [0.68,0.88]$ \cite{Farooq2017}. For more methods and models one can refer \cite{toribio,wang2006,Farooq2013}.

Following a similar line of thought, we use a model-dependent approach in constraining the transition redshift. In this paper, we use the updated compilation of 32 $H(z)$ data points obtained from cosmic chronometers and the Pantheon+ supernova dataset containing 1701 data points for the distance modulus. Further, we have used the MCMC technique to constrain the model parameters in the spatially flat and non-flat \lcdm and XCDM models. {\it This work improves upon earlier works by including the full covariance matrix for both datasets, which incorporates all statistical and systematic uncertainties }. We use the latest datasets and work with models that directly constrain the transition redshift instead of considering it a derived parameter. Additionally, we plot contours to study the correlations between different model parameters.
The paper is organized as follows: In section 2, we describe the \lcdm and XCDM models. The datasets used and the associated methodology is described in section 3. The final section discusses the results and conclusions of this work.

\section{Models}
In this paper, we have considered four different dark energy models. Using the fact that the second derivative of the scale factor $\ddot{a}=0$ at the transition epoch, we can derive a relation between the transition redshift and the relative densities of different components in the universe. Using this relation, we can find an equation for the Hubble Parameter in terms of $z_t$.

\subsection{\lcdm Model}

The acceleration equation in the \lcdm universe dominated by a constant density dark energy is given by:
\begin{equation}
     \frac{\ddot{a}}{a} = -\frac{4 \pi G}{3} \cdot (\rho_T + 3p_T)
\end{equation}

$\rho_T$ is the total energy density given by $\rho_T = \frac{\rho_{m0}}{a^3} + \rho_\Lambda$ and $p_T$ is the total pressure density.\\
Using the equation of state parameter $\omega = 0$ for the matter and $\omega = -1$ for dark energy in the acceleration equation:\\
\begin{equation}
    \frac{\ddot{a}}{a} = -\frac{4\pi G}{3} \left[ \rho_{m0}(1+z)^3 - 2\rho_\Lambda \right]
\end{equation}

Using the definition of the transition redshift with $\ddot{a} = 0$ and the equivalence of the energy densities to the normalized energy densities,  we obtain $z_t$ as :\\

\begin{equation}
    z_t = \bigg( \frac{2 \Omega_{\Lambda0}}{\Omega_{m0}} \bigg)^\frac{1}{3} -1
\end{equation}

Here $\Omega$ represents the normalized energy densities.
For the {\bf Flat \lcdm model}, the Hubble Parameter  is given as:

\begin{equation}
    H(z) = H_0 \left[ \Omega_{m0} (1+z)^3 + \Omega_{\Lambda0} \right]^\frac{1}{2}
\end{equation}
Substituting for $z_t$ along with  $ \Omega_{m0} + \, \Omega_{\Lambda 0} \, = \,1$, we obtain:
\begin{equation} 
    \boxed{H(z, f) = H_0 \left[ \frac{(1+z)^3}{\frac{1}{2}(1+z_t)^3 + 1} + \frac{(1+z_t)^3}{(1+z_t)^3 + 2} \right]^\frac{1}{2}} 
\end{equation}
Here, $f$ indicates the \textbf{free parameters} $H_0$ and $z_t$ in the Flat \lcdm model. 
\vskip 0.4 cm
Similarly for the {\bf Non-Flat \lcdm }model, the Hubble parameter is:
\begin{equation}
    H(z,f) = H_0 \left[ \Omega_{m0} (1+z)^3 + \Omega_{k0} (1+z)^2 + \Omega_{\Lambda0} \right]^\frac{1}{2}
\end{equation}
Where $\Omega_{k0}$ is a space curvature density parameter and $z_t$ for the Non-flat \lcdm model is now given as:

\begin{equation}
    z_t = \bigg( \frac{2 ( 1 - \, \Omega_{m0} -\Omega_{k0}}{\Omega_{m0}} \bigg)^\frac{1}{3} -1
\end{equation}
After substituting the value of $z_t$ and using the $\Omega_{m0} + \, \Omega_{\Lambda 0} + \, \Omega_{k0}\, = \,1 $, the Hubble parameter in Non-Flat \lcdm becomes:

\begin{equation}
    \boxed{H(z,f) = H_0 \left[ \frac{(1-\Omega_{k0})(1+z)^3}{\frac{1}{2}(1+z_t)^3 + 1} + \Omega_{k0}(1+z)^2 + \frac{(1-\Omega_{k0})(1+z_t)^3}{(1+z_t)^3+2} \right]^\frac{1}{2}}
\end{equation}
Here, $f$ indicates the \textbf{free parameters} $H_0$, $z_t$ and $\Omega_{k0}$.
\vskip 0.5 cm

\subsection{XCDM Model}
In the XCDM model, the dark energy acts as a dynamically evolving fluid. Here, the dark energy fluid pressure $p_X$ and energy density $\rho_X$ are related as:
\begin{equation} p_X = \omega_X\rho_X \end{equation}
where $\omega_X$ is the constant equation of state parameter having values less than $-\frac{1}{3}$.

Solutions to the fluid equation result in the energy density given as:
\begin{equation}
    \rho_X = \rho_{X0} \bigg( \frac{a_0}{a} \bigg)^{3(1+\omega_X)}
\end{equation} 

where the subscript "0" defines the current value of the parameters and thus $a_0$ is assumed to be unity. Substitution in the acceleration equation gives:
\begin{equation}
    \frac{\ddot{a}}{a} = -\frac{4\pi G}{3} \left[ \frac{\rho_{m0}}{a^3} + \rho_{X0} \bigg(\frac{1+3\omega_X}{a^{3(1+\omega_X)}} \bigg) \right]
\end{equation}

For the flat  XCDM model, the condition $\ddot{a} = 0$ results in:
\begin{equation} 
    z_t = \left[ \frac{-\Omega_{m0}}{\Omega_{X0} (1+3\omega_X) } \right]^{\frac{1}{3\omega_X}} -1 
\end{equation}
Here, $\Omega_{X0}$ is the normalized dark energy density.

For the {\bf Flat XCDM model}, the Hubble Parameter equation is:
\begin{equation}
    H(z,f) = H_0 \left[ \Omega_{m0} (1+z)^3 + \Omega_{X0}(1+z)^{3(1+\omega_X)} \right]^\frac{1}{2}
\end{equation}
Substituting for $z_t$, along with the condition $\Omega_{m0} + \, \Omega_{X0} \, = \,1$, we get:

\begin{equation} 
    \boxed{H(z, f) = H_0 \left[ \frac{(1+3\omega_X)(1+z_t)^{3\omega_X}(1+z)^3}{(1+3\omega_X)(1+z_t)^{3\omega_X}-1} + \frac{(1+z)^{3(1+\omega_X)}}{1-(1+3\omega_X)(1+z_t)^{3\omega_X}} \right]^\frac{1}{2}}
\end{equation}

Here, $f$ indicates \textbf{free parameters} $H_0$, $z_t$ and $\omega_{X}$.
\vskip 0.4 cm
For the {\bf Non-Flat XCDM model}, the Hubble Parameter can be written as:
\begin{equation} 
    H(z,f) = H_0 \left[ \Omega_{m0} (1+z)^3 + \Omega_{k0}(1+z)^2 + \Omega_{X0}(1+z)^{3(1+\omega_X)} \right]^\frac{1}{2}
\end{equation}

The transition redshift for this model can be written in term of the cosmological parameters as:
\begin{equation}
    z_t = \left[ \frac{-\Omega_{m0}}{ ( 1- \Omega_{m0}-\Omega_{k0})(1+3\omega_X) } \right]^{\frac{1}{3\omega_X}} -1
\end{equation}

By using the condition $ \Omega_{m0} + \, \Omega_{X0} \, +\Omega_{k0} = \,1 $  and substituting the value of  $z_t$ in the Hubble parameter equation, we obtain:

\begin{equation}
\boxed{ \resizebox{\linewidth}{!}{
 $H(z, f) = H_0 \left[ \frac{(1-\Omega_{k0})(1+3\omega_X)(1+z)^3(1+z_t)^{3\omega_X}}{(1+3\omega_X)(1+z_t)^{3\omega_X}-1} + \Omega_{k0}(1+z)^2 + \frac{(1-\Omega_{k0})(1+z)^{3(1+\omega_X)}}{1-(1+3\omega_X)(1+z_t)^{3\omega_X}}
\right]^\frac{1}{2}$ }}
\end{equation}
Here, $f$ indicates \textbf{free parameters} $H_0$, $z_t$, $\Omega_{k0}$ and $\omega_{X}$.

%% file: 2_data.tex
\section{Methodology and Data}

In this work, we use the updated 32 Hubble H(z) measurements obtained from passively evolving galaxies in the redshift range  $0.07<z<1.965$ and the 1701 distance modulus $\mu(z)$ measurements for Supernovae Type Ia in the redshift range $0.001<z<2.3$. We determine the best fit values of the parameters in different cosmological models by minimizing the combined $\chi^2$ for the two datasets which is given as:
\begin{equation}
    \chi^2_{total} = \chi^2_{CC} + \chi^2_{SNe}
\end{equation}
We use the publicly available {\sc emcee} \cite{emcee} python package to perform MCMC analysis using flat priors with ranges given in \textbf{Table \ref{tab:pr}}. The analysis gives the joint posterior probability distribution for the model parameters. The distribution is marginalized over other parameters to give an estimate for the maximum likelihood along with the 1$\sigma$ and 2$\sigma$ confidence intervals. Finally, we use the {\sc corner} \cite{corner} package to plot the 2D confidence contours.
The following section describes the observational data sets, statistical methods, and associated errors in detail.

\subsection{H(z) data}
The Hubble data was obtained from spectroscopic dating of massive, passively evolving low redshift $z \sim 2$ galaxies. Presently, these galaxies contain no active star-formation regions, with most of their stellar mass formed at $z > 1$. Chronometers are important as they measure the Hubble Parameter directly without assuming a particular cosmological model. Fundamentally, this technique determines the differential ages of adjacent pair of galaxies ($\Delta t$), given their differential redshift $\Delta z$. The ages of these galaxies are directly correlated to the metallicity of their stellar populations. This can be measured by the amplitude of the $4000 \textup{\AA}$ break in their absorption spectra \cite{Moresco_2016}. Finally, the Hubble function is given as:
\begin{equation} H(z) = -\frac{1}{1+z} \cdot \frac{\mathrm{d}z}{\mathrm{d}t} \end{equation}

To account for the complete set of systematic uncertainties, we include the full covariance matrix, represented as the sum of statistical and systematic uncertainties. The matrix is given as follows:
\begin{equation} Cov_{ij} = Cov_{ij}^{stat} + Cov_{ij}^{model} \end{equation}
where the systematic effects arise mainly due to the choice of different models used for estimating ages. The model covariance includes errors from the initial mass function (IMF), star formation history (SFH), stellar population synthesis (SPS) model, and stellar metallicity (SM).   
\begin{equation} Cov_{ij}^{model} = Cov_{ij}^{IMF} + Cov_{ij}^{SPS} + Cov_{ij}^{SFH} + Cov_{ij}^{SM} \end{equation}

To construct the covariance matrices we use the Mean Percentage Bias ($ \widehat{\eta^X}(z)$) table and the following relation from \cite{Moresco2020}.
\begin{equation} Cov_{ij}^{X} =  \widehat{\eta^X}(z_i) \cdot H(z_i) \cdot \widehat{\eta^X}(z_j) \cdot H(z_j) \end{equation}
Where $X$ represents the contribution from different error components. Using the 32 data points, we construct the $32\times32$ covariance matrix $Cov^{-1}_{stat+sys}$. We now calculate $\chi^2$ and the Likelihood as follows:
\begin{equation} -2\mathrm{ln}(\mathcal{L}) = \chi^2_{CC} = \boldsymbol{\Delta D^T} \cdot Cov^{-1}_{stat+sys} \cdot \boldsymbol{\Delta D} 
\end{equation}
where $\boldsymbol{\Delta D}$ is the residual vector defined as:
$ \boldsymbol{\Delta D_i} = H^{th}(z_i,\theta) - H^{obs}(z_i)$
, $\boldsymbol{\Delta D^T}$ represents its transpose and $\theta$ indicates the  model parameters. The $H^{th}$ denotes the Hubble parameter equation for the specific model while $H^{obs}$ is the observed value of the Hubble Parameter. 

\subsection{Supernova Data}
We use the latest Pantheon+ compilation, which analyses 1701 supernova light curves from 1550 distinct supernovae in the redshift range of 0.001 to 2.26. This data includes major contributions from CfA1-4, CSP, DES, PS1, SDSS and SNLS. The observed light curves were fitted using a SALT2 model, which returns the best fit value of the parameters $c$ (color), $x_1$ (stretch), and $x_0$ (overall amplitude) \cite{Scolnic2022}. 
Given the parameters, we can quantify $\mu_{obs}$, the observed distance modulus, using a linear model given as follows:
\begin{equation} \mu_{obs} = m_B + \alpha x_1 - \beta c - M - \delta_{\mu-bias} \end{equation}
 The nuisance parameters $\alpha$, $\beta$ and $M$ are jointly fitted with the cosmological parameters. Where  $\alpha$ and $\beta$ are the coefficients relating stretch and color to luminosity, $M$  is the absolute magnitude of the supernova and $\delta_{\mu-bias}$ represents the bias correction term. Now
\begin{equation} m_B \equiv -2.5\mathrm{log}_{10}(x_0) \end{equation}

\noindent Theoretically the distance modulus is given by:
\begin{equation} \mu_{th} = 5\mathrm{log}_{10} \bigg[\frac{D_L}{Mpc} \bigg] + 25 \end{equation}
where $D_L$, the luminosity distance is defined as:
\begin{equation} D_L (z) = (1+z) \cdot c \int_0^z \frac{dz'}{H(z')} \end{equation}
Here $c$ is the speed of light and 
$H(z)$ is the Hubble parameter equation for different models. 

Given $\mu_{obs}$ and $\mu_{th}$,   the residual is defined as:
\begin{equation} \boldsymbol{\Delta D_i}  = \mu_{th}(z_i,\theta) - \mu_{obs}(z_i) \end{equation}
where $\theta$ indicates the  model parameters.
The log-likelihood or $\chi^2$ relation can  now be written as:
\begin{equation} -2\mathrm{ln}(\mathcal{L}) = \chi^2_{SNe} = \boldsymbol{\Delta D^T} \cdot Cov^{-1}_{stat+sys} \cdot \boldsymbol{\Delta D}  \end{equation}

$Cov^{-1}_{stat+sys}$ is the $1701\times1701$ square covariance matrix as described in Brout et al. (2022) \cite{brout2022}.
Because there are 1701 light curves from 1550 SNe, the statistical covariance matrix includes the distance error ($\sigma_{\mu}^2$) as the diagonal entry and the measurement noise as the off diagonal terms for duplicate supernovae included in multiple surveys. This compilation improves upon earlier works by accounting for a much larger number of systematic uncertainties. These include errors from the measurement of redshift, peculiar velocities, and host galaxies; calibration of light curves and the SALT2 model fitting; extinction due to the Milky Way; and simulations of survey modeling, distance modulus uncertainty modeling, and intrinsic scatter models.

%% file: 3_results.tex
\section{Results}
\begin{figure}[ht]
    \centering
    \captionsetup{width=0.65\linewidth}
    \includegraphics[width = 0.65\linewidth]{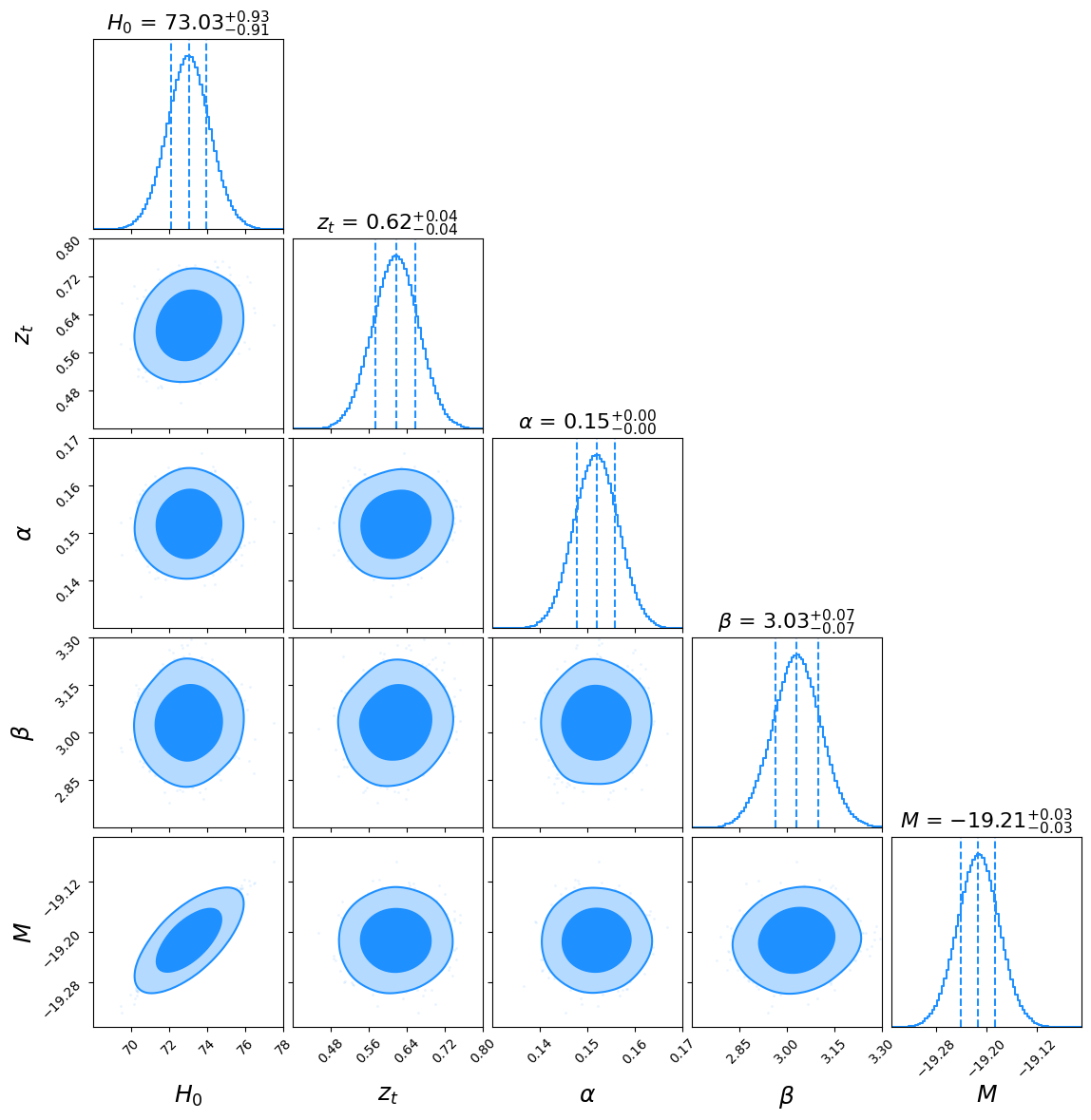}
    \caption{Joint Confidence Contours for the Flat \lcdm model with the CC + SNe dataset}
    \label{fig:lcdm}
\end{figure}

\begin{figure}[ht]
    \centering
    \captionsetup{width=0.65\linewidth}
    \includegraphics[width = 0.65\linewidth]{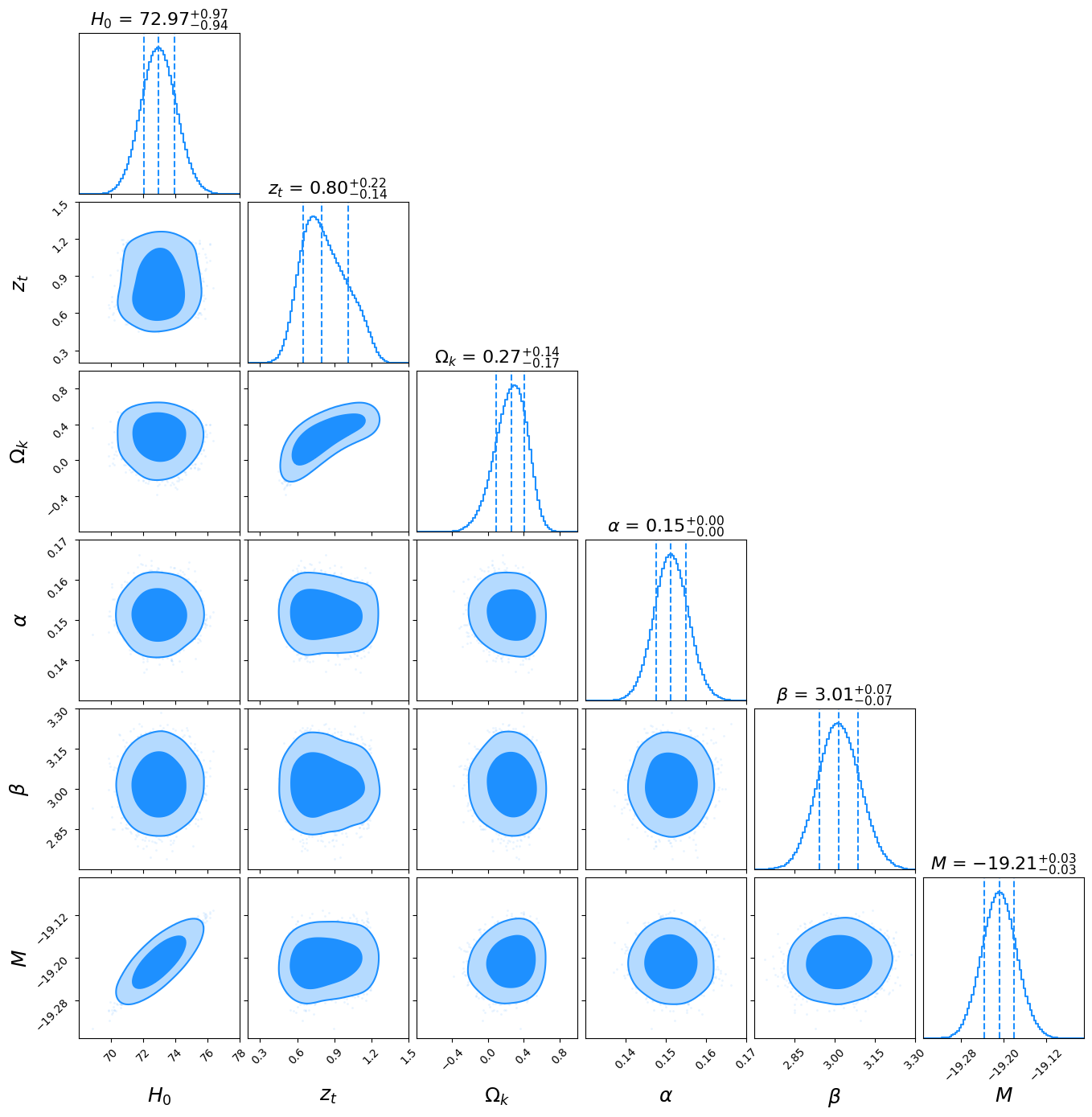}
    \caption{Joint Confidence Contours for the Non-Flat \lcdm model with the CC + SNe dataset}
    \label{fig:nlcdm}
\end{figure}

\begin{figure}[ht]
    \centering
    \captionsetup{width=0.65\linewidth}
    \includegraphics[width = 0.65\linewidth]{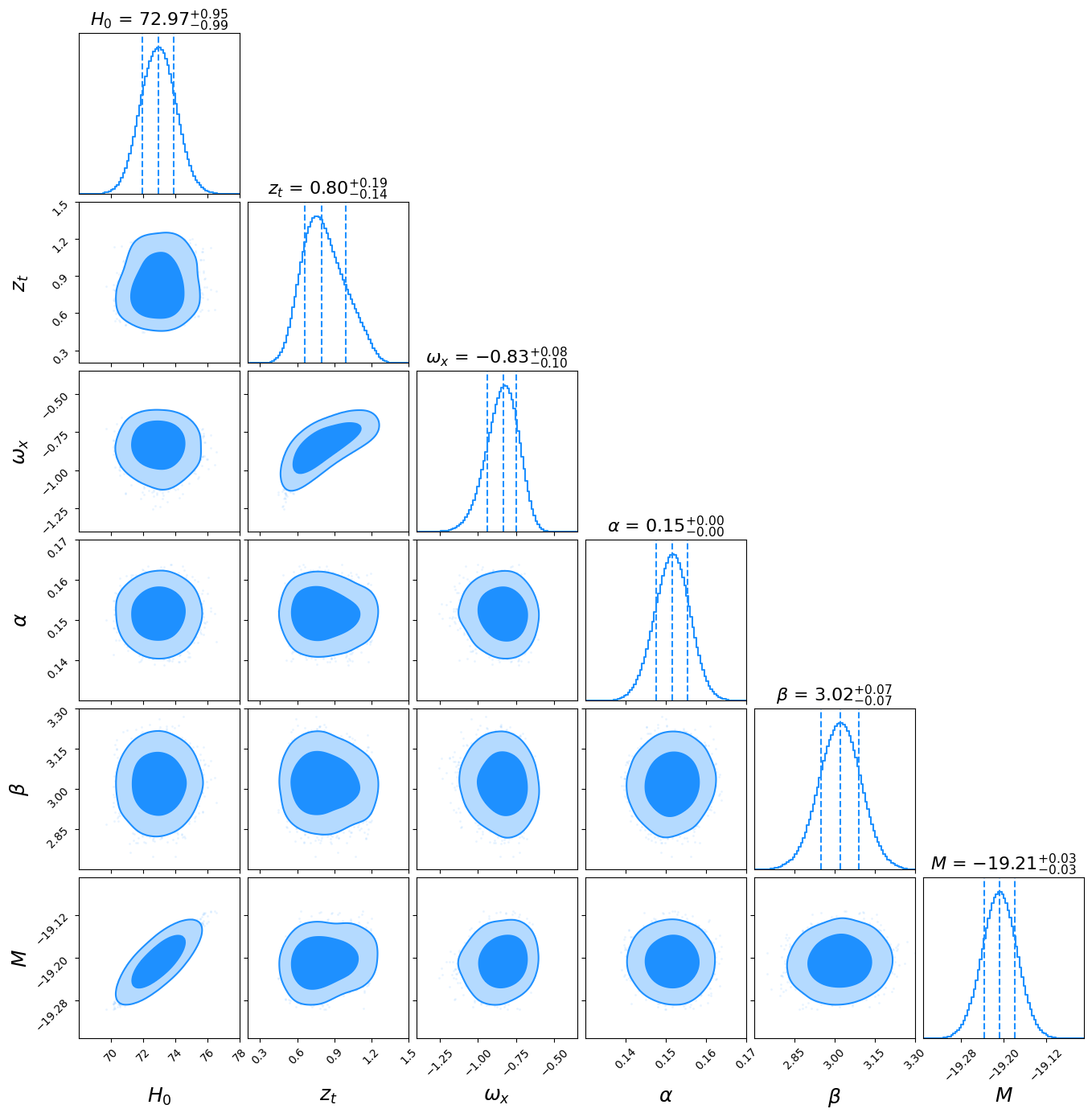}
    \caption{Joint Confidence Contours for the Flat XCDM model with the CC + SNe dataset}
    \label{fig:xcdm}
\end{figure}

\begin{figure}[ht]
    \centering
    \captionsetup{width=0.65\linewidth}
    \includegraphics[width = 0.65\linewidth]{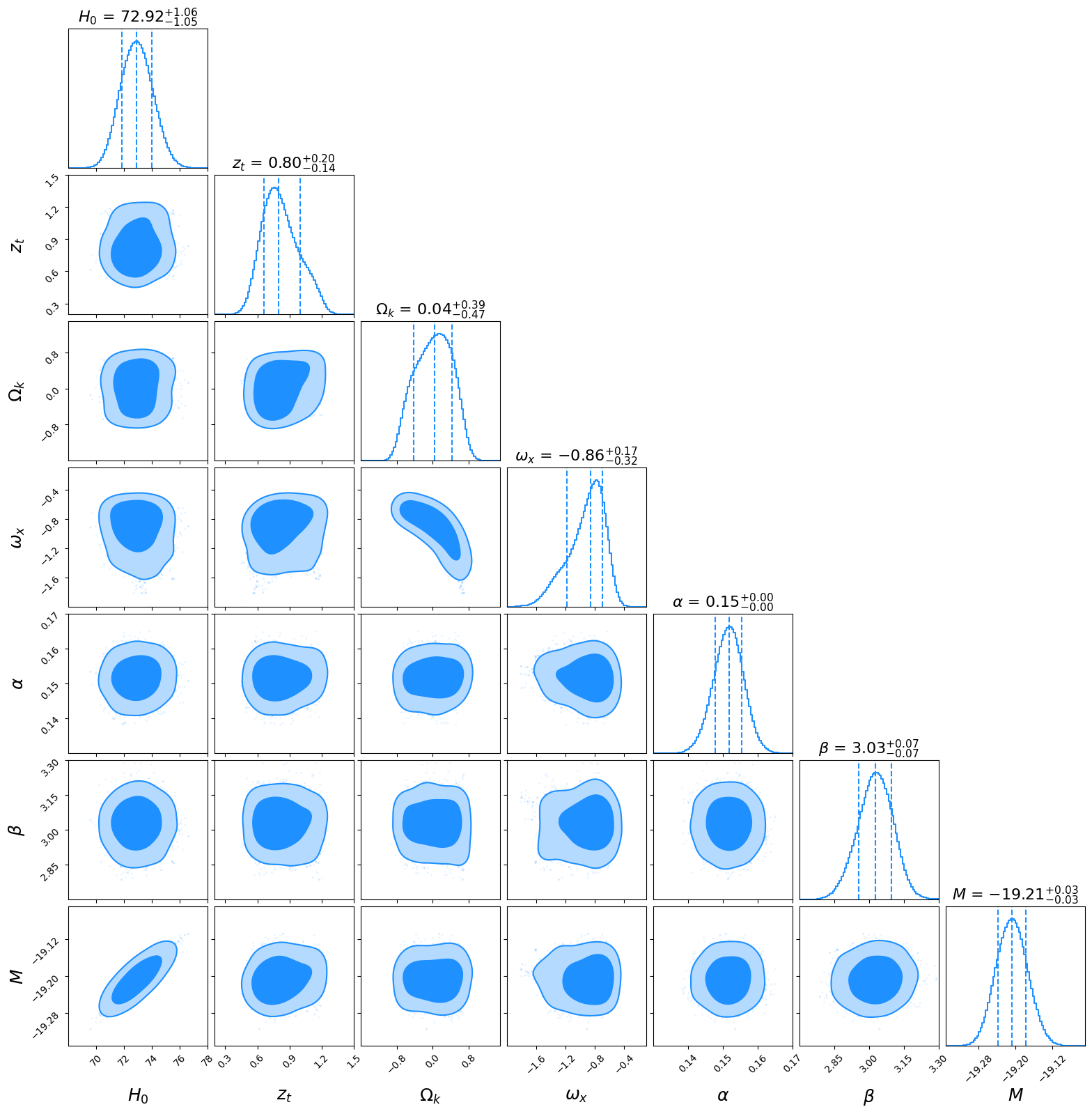}
    \caption{Joint Confidence Contours for the Non-Flat XCDM model with the CC + SNe dataset}
    \label{fig:nxcdm}
\end{figure}

In this paper, we use the updated available H(z)  and Supernovae  data-sets along with their full covariance matrices to obtain the constraints on the transition redshift, $H_0$, and other model parameters such as \,  $ \omega_X$ and $\Omega_{k0}$. Nuisance parameters $\alpha$,$\beta$ and $M$ are also jointly fitted to account for any additional bias. The 2-Dimensional contours and the 1-Dimensional posterior probability distributions for the cosmological parameter are shown in \textbf{Figure \ref{fig:lcdm}-\ref{fig:nxcdm}}. The best fit values of the model parameters obtained from different datasets are listed in \textbf{Table \ref{tab:BF}}.

\begin{itemize}
    \item For the \lcdm model the Hubble parameter $H_0$ and the transition redshift are tightly constrained. With both the datasets (SNe + CC), the spatially Flat model supports a $H_0= 73.034^{~0.937}_{-0.899}$ and a transition redshift of $z_t = 0.618^{~0.040}_{-0.042}$. 
    
    While the Non-Flat \lcdm model supports an open geometry ($\Omega_{k0} = 0.266^{~0.142}_{-0.171}$) with a $H_0$ value of $72.972^{~0.979}_{-0.933}$ and a transition redshift of $0.797^{~0.220}_{-0.144}$.
    
    \item The spatially Flat XCDM model suggest a dynamically evolving fluid ($\omega_X = -0.834^{~0.083}_{-0.101}$) with a $H_0 = 72.965^{~0.951}_{-0.981}$ and a transition redshift of $0.799^{~0.195}_{-0.140}$. 
    
    On the other hand, Non-Flat XCDM model also suggests slightly open geometry ($\Omega_{k0} = 0.044^{~0.389}_{-0.461}$) with a $H_0$ value of $72.922^{~1.071}_{-1.037}$, a transition redshift of $0.798^{~0.203}_{-0.135}$ and an equation of state  $ \omega_X = -0.863^{~0.167}_{-0.313}$. 

    \item The nuisance parameters are consistent across all models with little deviations between the Flat and the Non-Flat models. The parameter $\alpha$ ranges from 0.151 - 0.152, $\beta$ ranges from 3.014 - 3.030 and $M$ ranges from -19.207 - -19.213. 
\end{itemize}

\input{ms2022-0477table1}

%% file: ms2022-0477table1.tex
\begin{table}[h]
    \centering
    \captionsetup{width=0.45\linewidth}
    \begin{tabular}{c|c}
    \hline
    Parameter  &   Prior Range\\
    \hline
    $H_0$ & [50.0, 90.0] \\
    $z_t$ & [0.05, 1.2] \\
    $\Omega_{k0}$ & [-0.7, 0.7] \\
    $\omega_X$ & [-3.0, 0] \\
    $\alpha$ & [0.05, 0.2]\\
    $\beta$ & [2, 4]\\
    $M$ & [-19.5, -18.9]\\
    \hline
    \end{tabular}
    \caption{Flat priors assumed for the model and nuisance parameters}
    \label{tab:pr}
\end{table}

%% file: ms2022-0477table2.tex
\begin{table}
    \begin{adjustwidth}{2.5cm}{-11.5cm}
    \begin{adjustbox}{scale=1,center,angle=90}
    \resizebox{1.5\textwidth}{!}{
    \begin{tabular}{|M{0.15\linewidth}||M{0.13\linewidth}|c c c c c c c|}
        \hline
        &&&&&&&&\\
        \Large\textbf{Model} & \Large\textbf{Data Set} & \large$\boldsymbol{H_0}$ & \large$\boldsymbol{z_t}$ & \large$\boldsymbol{\Omega_{k0}}$ & \large$\boldsymbol{\omega_X}$ & \large$\boldsymbol{\alpha}$ & \large$\boldsymbol{\beta}$ & \large$\boldsymbol{M}$ \\
        &&&&&&&&\\
        \hline\hline
        \multirow{3}{5em}{\large \textbf{Flat} \lcdm} & SNe & \rule{0pt}{4ex} $73.500_{-0.978-2.035}^{~1.013~2.087}$ & $0.587_{-0.043-0.089}^{~0.043~0.096}$ & - & - & $0.152_{-0.004-0.008}^{~0.004~0.008}$ & $3.024_{-0.073-0.151}^{~0.072~0.149}$ & $-19.196_{-0.029-0.060}^{~0.028~0.060}$ \\ 
        & CC & \rule{0pt}{4ex} $67.841_{-5.604-11.337}^{~5.392~10.862}$ & $0.621_{-0.169-0.350}^{~0.163~0.338}$ & - & - & - & - & -\\
        & SNe + CC & \rule{0pt}{4ex} $73.034_{-0.899-1.983}^{~0.937~2.069}$ & $0.618_{-0.042-0.086}^{~0.040~0.083}$ & - & - & $0.152_{-0.004-0.008}^{~0.004~0.008}$ & $3.030_{-0.066-0.147}^{~0.068~0.144}$ & $-19.213_{-0.027-0.057}^{~0.027~0.057}$\\
        &&&&&&&&\\
        \hline
    
        \multirow{3}{5em}{\large \textbf{Non-Flat} \lcdm} & SNe & \rule{0pt}{4ex} $73.393_{-0.947-2.038}^{~1.000~2.078}$ & $0.723_{-0.160-0.254}^{~0.262~0.433}$ & $0.213_{-0.243-0.525}^{~0.179~0.280}$ & - & $0.151_{-0.004-0.008}^{~0.004~0.008}$ & $3.019_{-0.073-0.141}^{~0.075~0.152}$ & $-19.195_{-0.029-0.061}^{~0.029~0.060}$\\ 
        & CC & \rule{0pt}{4ex} $66.538_{-5.409-10.523}^{~5.526~11.782}$ & $0.608_{-0.194-0.452}^{~0.220~0.490}$ & $0.183_{-0.495-0.812}^{~0.362~0.494}$ & - & - & - & -\\
        & SNe + CC & \rule{0pt}{4ex} $72.972_{-0.933-1.958}^{~0.979~2.069}$ & $0.797_{-0.144-0.243}^{~0.220~0.373}$ & $0.266_{-0.171-0.372}^{~0.142~0.233}$ & - & $0.151_{-0.004-0.008}^{~0.004~0.008}$ & $3.014_{-0.069-0.145}^{~0.075~0.156}$ & $-19.208_{-0.028-0.057}^{~0.028~0.059}$\\
        &&&&&&&&\\
        \hline

        \multirow{3}{5em}{\large \textbf{Flat XCDM}} & SNe & \rule{0pt}{4ex} $73.361_{-1.015-2.074}^{~0.980~2.063}$ & $0.733_{-0.157-0.268}^{~0.223~0.415}$ & - & $-0.866_{-0.138-0.322}^{~0.106~0.171}$ & ${0.152_{-0.004-0.008}^{~0.004~0.008}}$ & ${3.016_{-0.074-0.146}^{~0.072~0.145}}$ & ${-19.195_{-0.030-0.062}^{~0.027~0.058}}$\\ 
        & CC & \rule{0pt}{4ex} $67.321_{-5.979-10.630}^{~8.644~17.790}$ & $0.566_{-0.147-0.395}^{~0.241~0.549}$ & - & $-1.005_{-0.784-1.645}^{~0.516~0.864}$ & - & - & - \\
        & SNe + CC & \rule{0pt}{4ex} $72.965_{-0.981-1.957}^{~0.951~1.972}$ & $0.799_{-0.140-0.242}^{~0.195~0.365}$ & - & $-0.834_{-0.101-0.218}^{~0.083~0.148}$ & ${0.152_{-0.004-0.008}^{~0.004~0.008}}$ & ${3.020_{-0.072-0.151}^{~0.071~0.149}}$ & ${-19.208_{-0.028-0.058}^{~0.029~0.058}}$\\
        &&&&&&&&\\
        \hline
        
        \multirow{3}{5em}{\large \textbf{Non-Flat XCDM}} & SNe & \rule{0pt}{4ex} $73.244_{-1.033-2.108}^{~1.078~2.190}$ & $0.784_{-0.170-0.307}^{~0.221~0.378}$ & $0.086_{-0.521-0.736}^{~0.397~0.555}$ & $-0.904_{-0.333-0.619}^{~0.199~0.288}$ & $0.151_{-0.004-0.008}^{~0.004~0.008}$ & $3.016_{-0.073-0.152}^{~0.071~0.139}$ & $-19.199_{-0.030-0.062}^{~0.031~0.067}$\\ 
        & CC & \rule{0pt}{4ex} $67.166_{-6.063-10.802}^{~8.008~18.167}$ & $0.600_{-0.207-0.462}^{~0.306~0.559}$ & $0.293_{-0.498-0.909}^{~0.290~0.392}$ & $-1.162_{-1.183-1.747}^{~0.668~1.023}$ & - & - & -\\
        & SNe + CC & \rule{0pt}{4ex} $72.922_{-1.037-1.959}^{~1.071~2.149}$ & $0.798_{-0.135-0.237}^{~0.203~0.368}$ & $0.044_{-0.461-0.698}^{~0.389~0.604}$ & $-0.863_{-0.313-0.625}^{~0.167~0.254}$ & $0.152_{-0.004-0.007}^{~0.004~0.008}$ & $3.028_{-0.072-0.150}^{~0.067~0.135}$ & $-19.207_{-0.031-0.058}^{~0.029~0.061}$\\
        &&&&&&&&\\
        \hline
    \end{tabular}
    }
    \end{adjustbox}
    \rotatebox{90}{\begin{minipage}{\textheight}
    \caption{ $1\sigma$ and  $2\sigma$  C.L. constraints on the model parameters}
    \label{tab:BF}
    \end{minipage}}
    \end{adjustwidth}
    
\end{table}

%% file: 4_conclusion.tex
\section{Discussion}
In this paper, we focus on constraining the transition redshift and build on previous works by using updated datasets with full covariance matrices and additional dark energy models. We express the Hubble parameter equation of dark energy models in term of $z_t$ and using the MCMC technique, obtain the contours between different model parameters. We observe that, compared to the H(z) data, the SNe data predicts an early time transition (except for the \lcdm model). Since we observe positive correlations between $z_t$ and other cosmological parameter ($\Omega_{k0}, \omega_X$) from the confidence contours, we can hypothesize that the exception of \lcdm model could be a consequence of these correlations. More research is needed to substantiate this claim, nonetheless all models support $z_t$ in the intermediate redshift range [0.61-0.79]. These results agree with past results obtained from other datasets and methodologies (mentioned in \textbf{Table \ref{tab:ztBF}}). We find negligible difference in the best fit values of  SNe Ia parameters in each dark energy model studied in this work.
Additionally, the constrained nuisance parameters are also consistent with the results obtained earlier in the literature \cite{SDSSres,chen2022}.

The obtained value of current Hubble Parameter ($H_0$) differ for the two datasets, further supporting the Hubble tension. The $H(z)$ data supports lower values of $H_0$ which is in concordance with the Planck CMB results while the Pantheon+ dataset support higher value of $H_0$ which again support the results  earlier obtained with the SNe dataset.\cite{planck,Riess2016}.

For all the Non-Flat models considered in the paper,  
the  non-flat \lcdm suggests a moderately open geometry ($\Omega_{k0} > 0$) but is still consistent with a spatially flat universe within $2\sigma$ limits. Similar observations of the curvature parameter were observed earlier by \cite{Farooq2017,yang2020}. The Non-Flat XCDM model, on the other hand, indicates a very mild deviation from a Flat Universe but has larger error bounds on the curvature density of the universe.

For the dynamical dark energy models, there is mild variation in the equation of state parameter ($\omega_X \neq -1$). Nonetheless, the \lcdm model ($\omega_X=-1$) can be easily recovered within $2\sigma$ levels. Our results are consistent with those obtained recently with the Pantheon+ compilation \cite{brout2022} and the 2019 DES Compliation \cite{DES2019}.
\vskip 0.1cm
As mentioned above, the Non-Flat models support an open geometry, although, the Non-Flat \lcdm model indicates a much stronger positive curvature ($\Omega_{k0}=0$ is 2$\sigma$ away) as compared to the XCDM model ($\Omega_{k0}=0$ is 1$\sigma$ away).
This shows that, when the equation of state is allowed to vary, a flat universe is more statistically probable. Thus, a strong negative correlation exists between the dark energy equation of state and the curvature density which can also be seen in confidence contours for the Non-flat XCDM model (\textbf{Figure \ref{fig:nxcdm}}). This degeneracy is further discussed in \cite{Clarkson2007,Ichi_2006} which explore models with different assumptions and discuss the importance of constraining dark energy models in association with the curvature. They also mention the implications of assuming zero curvature on the equation of state parameter. More information on this degeneracy can be found in \cite{deg1,deg2}.

Finally, we observe that using the combined, updated  datasets of H(z) and SNe along with their full covariance matrices, the best fit value of transition redshift lies in the range $0.618 < z_t < 0.799$ for all four dark energy models with the standard Flat \lcdm model having the lowest error bars compared to other models. These results are in general agreement with past analyses and the Planck's results within $ 2\sigma$ level.

%% file: ms2022-0477table3.tex
\begin{table}[h]
    \centering
    \begin{adjustbox}{scale=1,center}
    \resizebox{\linewidth}{!}{
    \begin{tabular}{M{2cm}||p{3.7cm}||M{6cm}|M{3cm}|M{3.5cm}}
        \hline
        &&&&\\
        \large\textbf{Method} & \large\textbf{Models} & \large\textbf{Data Set} & $\mathlarger{\mathlarger{\mathlarger{\boldsymbol{z_t}}}}$ & \large\textbf{References} \\
        &&&&\\
        \hline\hline
        \multirow{3}{7em}{Likelihood Maximization} & \rule{0pt}{3.5ex}KM: q($z$) & SNe(SNLS) & 0.61 & \cite{Cunha2008} \\
        & KM: $\omega(z)$ & \BTstrut BAO + CMB(WMAP) + SNe(Union) & \BTstrut 0.7 - 1 & \BTstrut \cite{wzparam}\\
        & KM: q($z$) & \BTstrut Age of Galaxies + Strong Lensing + SNe(JLA) &  0.6 - 0.98 & \cite{Rani_2015}\\
        & \rule{0pt}{3.5ex}KM: H($z$), $D_C(z)$, q($z$) & \Tstrut CC + SNe(JLA) & \Tstrut 0.806 - 0.973 & \Tstrut \cite{Jesus_2018}\\
        & KM: q($z$) & \BTstrut CC + BAO + SNe(Pantheon) + CMB & \BTstrut 0.593 - 0.792 & \BTstrut \cite{abdulla}\Bstrut\\
        \cline{2-5}
        
        & \Tstrut \multirow{3}{6.5em}{\lcdm Model} & \Tstrut CC & \Tstrut 0.64 & \Tstrut \cite{Moresco_2016}\\
        & & \Tstrut CC + BAO & \Tstrut 0.723 - 0.832 & \Tstrut \cite{Farooq2017,Farooq2013} \\
        & & \Tstrut CC + BAO + SNe(Pantheon) & \Tstrut 0.69 & \Tstrut \cite{toribio} \\
        & & \BTstrut \textbf{CC + SNe (Pantheon+)} & \BTstrut \textbf{0.61 - 0.82} & \BTstrut \textbf{Present Work}\Bstrut\\
        \cline{2-5}

        & \Tstrut CF: H($z$) & \Tstrut CC + BAO + SNe(Pantheon) & \Tstrut 0.6857 & \cite{Koussour2022}\\
        & CF: H($z$), q($z$), j($z$) & \Tstrut CC + BAO + SNe(Union) & \Tstrut 0.77 - 0.86 & \Tstrut \cite{capo2014}\\
        & CF: a($t$) & \Tstrut BAO + SNe(Union2.1) & \Tstrut 0.28 - 0.63 & \Tstrut \cite{Mu2016}\\
        & CF: H($z$), q($z$) & \Tstrut SNe (Pantheon) + BAO + GRB & \Tstrut 0.739 - 0.831 & \Tstrut \cite{muccino2022}\Bstrut\\
        \cline{2-5}

        & \rule{0pt}{3.5ex} \multirow{3}{6.5em}{XCDM Model} & \Tstrut CC + BAO & \Tstrut 0.684 - 0.813 & \cite{Farooq2017,Farooq2013} \\
        & & \BTstrut \textbf{CC + SNe(Pantheon+)} & \BTstrut \textbf{0.77 - 0.79} & \BTstrut \textbf{Present Work}\Bstrut\\
        \cline{2-5}

        & \Tstrut $\phi$CDM Model\Bstrut & \Tstrut CC + BAO & \Tstrut 0.690 - 0.885 & \Tstrut \cite{Farooq2017,Farooq2013}\\
        \cline{2-5}
        \hline

        \multirow{4}{6.5em}{Regression} & \BTstrut GP: H($z$), $D_L(z)$ & \Tstrut CC + SNe(Pantheon) & \Tstrut 0.57 - 0.69 & \Tstrut\cite{Jesus_2020}\\ 
        & GP: H($z$), q($z$) & \Tstrut CC + BAO & \Tstrut 0.637 - 0.71 & \Tstrut \cite{Toribio_2021}\\
        & GP: q($z$) & \Tstrut CC + SNe(Pantheon) & \Tstrut 0.61 & \Tstrut \cite{Yang_2020}\\
        & \Tstrut GP: H($z$) & \Tstrut CC + BAO & \Tstrut 0.44 - 0.65 & \Tstrut \cite{Yu2018}\Bstrut\\
        \cline{2-5}
        & \Tstrut WFR: q($z$), j($z$) & \Tstrut CC + BAO + SNe(Pantheon + MCT)\Bstrut & \Tstrut 0.8 & \Tstrut \cite{GM2019}\\
        \cline{2-5}
        & \Tstrut LOESS+SIMEX & \Tstrut Age of Galaxies + Strong Lensing + SNe(JLA) \Bstrut & \Tstrut 0.7 & \Tstrut \cite{Rani_2015}\\
        \hline

    \end{tabular}
    }
    \end{adjustbox}
    \caption{A summary of the current constraint on the transition redshift obtained from different works. (KM: Kinematic Models, CF: Cosmographic Functions, GP: Gaussian Process, WFR: Weighted Function Regression)}
    \label{tab:ztBF}
\end{table}